# Absolute surface energy determination

J.J.Métois, P.Müller[1]

Centre de Recherche sur les Mécanismes de la Croissance Cristalline[2]
Campus de Luminy, case 913, F-13288 Marseille Cedex 9, France

**ABSTRACT**

Experimental determination of absolute surface energies remains a challenge. We propose a simple method based on two independent measurements on 3D and 2D equilibrium shapes completed by the analysis of the thermal fluctuation of an isolated step. Using then basic equations (Wulff' theorem, Gibbs-Thomson equation, thermodynamics fluctuation of an isolated step) allows us to extract the absolute surface free energy of a singular face. The so-proposed method can be applied when (i) all orientations exist on the equilibrium shape, (ii) the surface stress is isotropic. This procedure is applied to the case of Si(111) where we find $0.59 Jm^{-2} \leq \gamma_{(111)} \leq 0.83 Jm^{-2}$ at 1373 K.

## Introduction

Since the pioneering works of Curie [1] and Wulff [2] it is well known that the equilibrium shape of a free crystal is the shape that minimises its surface free energy. More precisely, the Wulff theorem written under the form

$$\frac{\gamma_i}{h_i} = \lambda \qquad (1)$$

allows to measure the relative surface free energies $\gamma_i/\gamma_j$ of two facets i and j as the ratio of the two central distances $h_i/h_j$ of the facets i and j to the central Wulff point W (see figure 1a).

For a deposited crystal (but free of any elastic relaxation [3-5]) the adhesion energy $\beta$ between the deposited crystal and its underlying substrate also occurs via an additional equation valid for the face parallel to the interface. According to the so-obtained Wulff

---

Kaishev theorem [3-5], the equilibrium shape of the free crystal is preserved since equation (1) is still valid, but the crystal is truncated by the substrate, the truncature being dependent upon the adhesion energy.

Wulff or Wulff-Kaishew theorem has thus been used to study the surface free energy anisotropy (known as $\gamma$-plot) of various organic [6-8] metallic [9-13] and more recently semi-conductor [14-17] materials. Nevertheless the experimental determination of absolute surface free energies of crystalline faces remains a challenge[3].

To this aim several experimental approaches have been proposed recently [18-21].

The Bonzel approach [18] is based on the measurement of the temperature dependence of the three dimensional (3D) equilibrium shape. More precisely, (i) the experimental temperature dependence of the facet shape and size is combined with an analytical expression (Ising model) of the temperature dependence of the step free energy. The absolute step energy is thus obtained provided the facet extension at zero temperature $x_F(0)$ may be estimated. (ii) For regular 3D shape, the surface free energy is then simply obtained by scaling the step free energy by the ratio $H_F(T)/x_F(T)$, where $H_F(T)$ is the central distance of the facet to the Wulff point. Though this experimental approach is model-dependent (Ising model) and needs an estimation of the value of $x_F(0)$, it leads to values of the surface free energies in quite a good agreement with theoretical calculations (though theoretical data belongs to a quite spread spectrum of values!).

Measurements of absolute step free energies have also been recently proposed on the basis of a detailed study of the 2D equilibrium shape and of its thermal fluctuations [22-27]. The new approach we propose for the determination of the absolute surface energy determination proceeds from both previous methods since it requires a careful analysis of the 3D, and 2D equilibrium shapes coupled to a statistical analysis of the thermal fluctuations of an isolated step. Contrary to the Bonzel approach, the so-obtained results are not model dependent and do not need any estimation of $x_F(0)$.

This method will be first derived from well-known classical results, and then illustrated in the case of the Si(111) surface beyond the $(7x7) \leftrightarrow (1x1)$ transition temperature.

# I/ Experimental determination of surface energy: the principle

The surface free energy determination is based on three interconnected experiments. The first one consists in the determination of the size of a facet i on the 3D equilibrium shape that depends upon the ratio $\beta_i/\gamma_i$. The second experiment is based on a careful study of the 2D equilibrium shape of an island which gives access to the ratio $\beta/\tilde{\beta}$ ($\tilde{\beta}$: step stiffness) through 2D Wulff's theorem and Gibbs-Thomson relation.. At last a statistical analysis of the thermal fluctuation of an isolated step gives access to the absolute value of the step stiffness $\tilde{\beta}$. The simultaneous use of these three experimental data leads to the experimental determination of the absolute surface free energy $\gamma_i$. From the experimental point of view one has just to be able to perform true equilibrium experiments that means (at last at high temperature) to be able to compensate the desorbing flux by an impinging one as described by Heyraud et al. [28].

## I.1/ Size of a facet on the 3D equilibrium shape: determination of the ratio surface to step energy $\beta_i/\gamma_i$

Since Landau [29], the equilibrium shape of a 3D crystal can be expressed as the pedal of the anisotropy of the surface free energy. More precisely for a cylindrical non-singular face (tangentially joined to the curved part) the equilibrium shape is given by:

$$\lambda z = \sigma - p\frac{\partial \sigma}{\partial p} \qquad (2a)$$

$$\lambda x = \frac{1}{a}\frac{\partial \sigma}{\partial p} \qquad (2b)$$

where the super-saturation, $\lambda$, is a constant scaling the equilibrium shape, p is the slope of the vicinal surface (terraces + steps), $\sigma = \gamma\sqrt{1+p^2}$ is the projected free surface energy on the terrace plane and finally a is the thickness of the surface layer.

For a singular face, the surface energy cannot be analytically developed so that equations (2) are of no use. Nevertheless the singular face being connected to an adjacent vicinal face on the equilibrium shape (see figure 1b), equations (2) are valid at the edge where the vicinal face and the flat (singular) one meet together. The quantity

---

[3] Recommended values of surface free energies are generally obtained from zero creep or pendant drop for liquid or cleavage experiment for crystal. Nevertheless cleavage experiment can only give access to absolute surface energies of cleavable faces !!

$$\lambda x_F = \frac{1}{a}\frac{\partial \sigma}{\partial p}\bigg|_{p=0} \qquad (3)$$

thus defines the lateral extension $x_F$ of the facet (see figure 1b).

Using now the surface energy development valid for a vicinal face, $\sigma(p) = \gamma + \beta p + \phi p^3$ where $\gamma$ is the surface free energy of the nominal flat surface, $\beta$ is the step free energy and $\phi$ is the step-step interaction (for example see [30]), the size of the flat facet reads:

$$x_F = \frac{1}{\lambda}\frac{\beta}{a} \qquad (4)$$

The size of the facet is thus determined by the ratio $\beta/\lambda$. However since the Wulff theorem (1) is valid the scaling factor $\lambda$ can be suppressed so that the lateral extension of the facet i now reads:

$$x_F = \frac{\beta_i}{\gamma_i}\frac{H_i}{a} \qquad (5)$$

Thus $x_F$ being a measurable quantity, the ratio $\beta_i/\gamma_i$ can be simply obtained by measuring the lateral size of the facet i on the 3D equilibrium shape.

**I.2/ Study of the 2D equilibrium shape: determination of the ratio step energy to step stiffness $\beta_i/\widetilde{\beta}_i$**

Let us now recall other well-known results about the equilibrium shape of 2D supported islands for which the Wulff theorem (1) now reads:

$$\frac{\beta_i}{h_i} = \lambda' \qquad (6)$$

where $\lambda'$ is the 2D supersaturation and where the step free energy $\beta_i$ replaces the surface free energy $\gamma_i$.

As in the previous section, equation (6) is useful only if the scaling factor $\lambda'$ is known. This is indeed the case as can be seen from the Gibbs-Thomson equation connecting the step stiffness $\widetilde{\beta}_i$ to the radius of curvature of the step by means of the scaling factor $\lambda'$ expressed as [31] (see figure 2) :

$$\frac{\widetilde{\beta}_i}{R} = \lambda' \qquad (7)$$

Injecting (7) in (6) thus leads to the following relation:

$$\frac{\beta_i}{\tilde{\beta}_i} = \frac{h_i}{R} \qquad (8)$$

Once more, $h_i/R$ being a measurable quantity, one can then determine $\beta_i$ provided that $\tilde{\beta}_i$ is available from another experiment.

**I.3/ Thermal fluctuation of an isolated step: determination of the step stiffness $\tilde{\beta}_i$**

Let us consider an isolated step, i.e a step for which fluctuations are not limited by nearest neighbours steps. The step stiffness $\tilde{\beta}$ acts against the fluctuation of position of the step. For a step (length L) pinned between two anchored defects A and B (see figure 3), the average fluctuation can easily be calculated from elemental statistical physics of a string. More precisely by application of the equipartition theorem, the mean square fluctuations averaged over time and the length exactly reads [32,34]:

$$\langle x^2 \rangle = \frac{kTL}{6\tilde{\beta}} \qquad (9)$$

Note that such fluctuation analysis have been performed on various systems in order to obtain the step stiffness $\tilde{\beta}$ (for example see [33,35]).

**I.4/ Partial conclusion**

As a partial conclusion : **(i)** the analysis of the thermal fluctuations of an isolated step gives access to $\tilde{\beta}_i$ (see (9)), **(ii)** a further careful study of the 2D equilibrium shape gives access to the step free energy $\beta_i$ (via the ratio $\beta_i/\tilde{\beta}_i$, see equations (8-9)), **(iii)** using the previous points (i) and (ii) the lateral size of a flat facet i on the 3D equilibrium shape leads to the absolute value of the surface energy of the singular face $\gamma_i$ via the ratio $\beta_i/\gamma_i$.

Let us underline that such an absolute determination of the surface energy of a singular facet is only based on very simple and fundamental well-known equations and does not need any peculiar assumption on the temperature-dependence of the size of the facet. The procedure is thus quite general and can be applied each time **(i)** the 2D and 3D equilibrium shapes (with all orientations present) are available, **(ii)** the geometrical properties of the 2D and 3D equilibrium shapes as well as the thermal fluctuations of an isolated step can be measured with a sufficient accuracy, **(iii)** the flat surface is characterised by an isotropic stress so that the 2D equilibrium shape (6) is not modified by surface stress (as clearly shown in [36] for Ge(001) surface). It is the case of Si (111) surface above the $7x7 \leftrightarrow 1x1$ transition, for which the recovered three-fold symmetry provides an isotropic surface stress and all data are available from previous studies [16-17, 28,34].

## II/ Application to the case of Si(111) surface

### II.1/ Experimental set-up

The 3D equilibrium shape (ES) is reached by heating a Si column (obtained by etching) to roughly 1623 K. The heating is interrupted when the equilibrium shape is reached at the apex of the so formed bulb (for more details see [16,17]). The ES is then observed in-situ at 1373 K by ultrahigh vacuum transmission microscopy, and after quenching ex situ by conventional scanning electron microscope. A cross section of such a bulb is shown on figure 4a in the <110> direction. Let us point out that several experimental points confirm that the 3D equilibrium state has been reached: (i) the stable shape profile is time and size independent, (ii) a reversible shape change may be produced by weak temperature variation, (iii) the symmetrical axis of the 3D crystal is not the macroscopic symmetrical axis of the initial column (for more details see [16,17]) but the crystallographic one.

Both 2D equilibrium shape and the thermodynamics step fluctuations are observed by in-situ reflection electron microscopy at T=1373 K (residual pressure $P \approx 10^{-9} Torr$ during the experiment). At such a temperature, Silicon evaporation occurs so that the equilibrium condition can only be reached if this evaporation flux is exactly compensated by an incoming flux due to another Si sample. To do that, two Silicon samples are faced one to the other and the buffer sample is heated independently of the studied sample (for more details see [37,38]). It is thus possible to perfectly control the equilibrium state of the sample. More precisely, for too low (respectively too great) incoming flux, the steps recede (respectively advance) while at equilibrium they only fluctuate around their position.

At last, one has to keep in mind that, due to the grazing incidence, the REM image is shortened in the direction perpendicular to the electron beam. The reduction factor usually reported is around 49±2 when observations are performed under the same conditions [39]. In the following we use the mean value of 49. However the poor image definition does not allow to discriminate between 48 and 50.

**II.2/ Experimental results**

<u>II.2.1/ 3D equilibrium shape: Experimental determination of $\beta/\gamma$</u>

On figure 4a is reported the 3D equilibrium shape of a Silicon bulb obtained from [16-17]. Let us consider the (111) facet and more precisely its lateral extension (see figure 4b). It can be noted that the left and right lateral extensions of the (111) facet (with respect to the central axis z normal to the (111) face and passing through the Wulff point) are not equal, which is quite normal when considering the asymmetry of the two neighbouring vicinal faces. Indeed in the $x$ direction (see figure 4b) the rounded part connects the (111) face to another (111) face whereas in the $\bar{x}$ direction, the rounded part connects the (111) face to a (001) face. Thus in the $x$ direction the vicinal face is characterised by <110> steps with at the edge a {111} micro facet, whereas in the $\bar{x}$ direction the vicinal face still involves <110> steps but now with at the edge a (001) micro facet (see figure 5). Therefore the two edges of the flat (111) face are not equivalent and do not present the same step free energy, so that the lateral extension of the (111) facet under consideration is not symmetric with respect to the z axis. Since the equilibrium shape is only reached at the apex of the bulb [17], we measure the lateral extension of the top (111) facet and which leads, using (5), for the two considered steps:

$$3.48\ 10^{-11} m \leq \frac{\beta_{<110>}^{(111)\to(111)}}{\gamma_{(111)}} \leq 3.82\ 10^{-11} m \quad \text{and} \quad 4.44\ 10^{-11} m \leq \frac{\beta_{<110>}^{(111)\to(001)}}{\gamma_{(111)}} \leq 4.77\ 10^{-11} m \qquad (10)$$

Such a determination gives access to the angular dependence of the step energy on a Si(111) surface. Indeed, three-fold symmetry implies that $\beta(\theta) = \beta_o + \beta_1 \cos 3\theta$ with $\beta_o \approx 4.12$ and $0.31 \leq \beta_1 \leq 0.64$ (scale factor $10^{11} \gamma_{(111)}$). The so-obtained step energy anisotropy is reported in figure 6.

<u>II.2.2/ 2D equilibrium shape: Experimental determination of $\beta/\widetilde{\beta}$</u>

The experimental determination of $\beta/\widetilde{\beta}$ requires a careful study of the equilibrium shape of a 2D island on a Si(111) surface. This has been done by Scanning Electron Microscopy (SEM) [40] or by REM [37,38] on 2D islands and by Atomic Force Microscopy (AFM) on 2D pits [41]. Furthermore, the 2D equilibrium shape can also be directly calculated as the pedal of the $\beta$-plot. This is illustrated in figure 2 where is plotted the 2D equilibrium shape calculated for $\beta_o \approx 4.12$ and $\beta_1 = 0.31$, i.e for the weaker anisotropy compatible with the experimental data obtained in II.2.1 (in the same figure is also reported the inward circle giving the experimental ratio $\beta/\widetilde{\beta}=h/R=0.62$ [4]). Taking into account all the experimental incertitude of (10), one finds that $0.45 \leq \beta/\widetilde{\beta} \leq 0.62$.

This large incertitude may be reduced by the study of the true experimental 2D equilibrium shape. Some SEM [40] and AFM images [41] reveal near circular islands leading to isotropic step energy in contradiction with our experiments on 3D equilibrium shape. However these images do not correspond to true equilibrium conditions since the evaporating flux is not compensated by some additional impinging flux. On one other hand, in our REM experiment, where true equilibrium conditions are reached, the image distortion is a serious limitation for an accurate determination of the equilibrium shape. More precisely restoring the true equilibrium shape requires for us to extend the recorded image by a factor 49 in one direction which means that the initial pixel size becomes the main source of incertitude. Nevertheless, though its boundary definition is quite large, the so reconstructed image given in figure 7b (negative image, the wide edge of the 2D island being in white) is quite well fitted by the equilibrium shape calculated for $\beta/\widetilde{\beta}=h/R=0.62$, namely for the weaker anisotropy compatible with 3D equilibrium shape experiment

$$\beta_{<110>}^{(111)\rightarrow(111)}/\gamma_{(111)} \approx 3.82\ 10^{-11} m \quad \text{and} \quad \beta_{<110>}^{(111)\rightarrow(001)}/\gamma_{(111)} \approx 4.44\ 10^{-11} m \qquad (11)$$

Therefore, in the following we will use the value:

$$\beta/\widetilde{\beta}=h/R=0.62 \qquad (12)$$

### II.2.3/ Experimental determination of $\widetilde{\beta}$

At last the step stiffness extracted from the thermal fluctuation of an isolated step pinned between two anchored defect and measured at 1173 K is found to be $\widetilde{\beta}=7.3 10^{-11} Jm^{-1}$.[34] . Nevertheless, at 1373 K, the step stiffness determined by analysis of roughly 5000 frames only reaches 50% of this value, as confirmed by analysis of temporal correlation of step wandering [35]. Thus in the following we will take at 1373 K

$$3.65.10^{-11} Jm^{-1} \leq \widetilde{\beta} \leq 5.10.10^{-11} Jm^{-1} \qquad (13)$$

---

[4] We consider obviously that the 2D equilibrium shape is limited by the steps having the weaker step energy that means $\beta=\beta_{<110>}^{(111)\rightarrow(111)}$

II.2.4/ Discussion

All these experimental data lead to $0.59 Jm^{-2} \leq \gamma_{(111)} \leq 0.83 Jm^{-2}$. This value can be compared to other data given in the literature (see table I). They all come from theoretical calculations using more or less complex approximations [42,43]. Since measured at higher temperature, our experimental value is lower than the ones calculated at lower temperature. It is also interesting to compare our experimental value to the one measured at the melting point (1688 K) by the sessile droplet method [44] : $\gamma = 0.825 Jm^{-2}$.

| Sander et al [42] 0K | Sander et al [42] 77 K | Pluis et al. (RT) [43] | Pluis et al 350K [43] | This work 1373 K | Huang liquid 1688 K [44] |
|---|---|---|---|---|---|
| First principles calculations | Cleavage experiments | Thermodynamic calculation | Thermodynamic calculation | Absolute measurement | Absolute measurement by sessile drop method |
| 1.82 | 1.24 | 1.03 | 1.2 | 0.59–0.82 | 0.825 |

Thus when considering the temperature effect that lowers the surface energy, our result is quite consistent both with calculated data and experimental data near the melting point. Nevertheless the value we obtain seems a bit weak even though one keeps in mind the poor accuracy due to the resolution of our camera.

III/ Conclusion:

The new method we propose for measuring absolute surface energies implies to connect three different experiments and very basic surface thermodynamics. It can be applied when all the orientations exist on the equilibrium shape and when the surface stress is isotropic. In case of (111) surface, in spite of the poor accuracy of the REM image, all experimental results are in a good agreement . More precisely the step free anisotropy obtained from the 3D equilibrium shape is compatible with the 2D equilibrium shape, but not with STM and AFM data obtained out of equilibrium (i.e without compensating flux). The so-obtained free energy is a bit weaker than expected. It could be due to the high adatom density measured in equilibrium condition by STM [45] (the adatom density is 10-11 % higher than on the bulk-terminated surface) or by Reflection Electron Microscopy [46] ($\theta \approx 0.2$ on the Si(111) surface) which can lower the free surface energy by 20%. A study of equilibrium shape change with temperature could be helpful to understand lowering effect.

Figure captions :

**Figure 1 :** Definition of the geometrical data: **a)** Wulff construction of a polyedrical crystal. The central distances $H_i$ measured from the Wulff point W characterise the surfaces i of surface free energy $\gamma_i$, **b)** definition of the lateral extension of a facet.

**Figure 2:** Equilibrium shape of a 2D crystal (thick curve) obtained as the pedal of the step free energy $\beta(\theta)=\beta_0+\beta_1\cos3\theta$ (with $\beta_0$=4.12 and $\beta_1$=0.31), defining once again the central distance h to the Wulff point W. On the figure is also plotted the circle defining the radius of curvature R of the step. Here $h/R$=0.62.

**Figure 3:** REM image of an isolated step fluctuating at T=1373 K

**Figure 4:** 3D equilibrium shape ([110] zone) . (a) SEM image of a Silicon bubble the apex of which is the equilibrium shape at 1373 K, (b) Sketch of the SEM image defining the Wulff point (W), the facet orientation, the central distances and the lateral extension of the (111) facets (arrows). The lateral extension measurements have been made on the top (111) surface where the equilibrium shape is not perturbed by the neck of the bulb (for more details see [15,16])

**Figure 5:** Si(111) terrace with two limiting steps. It is easy to see that there are two kinds of <110> steps with {111} or {001} microfacets.

**Figure 6:** $\beta$-plot obtained for $\beta(\theta)=\beta_0+\beta_1\cos3\theta$ (with $\beta_0$=4.12 and $\beta_1$=0.31)

**Figure 7:** **(a)** 2D equilibrium shape obtained by REM at 1373 K, **(b)** true 2D equilibrium shape obtained after correction by the factor 49. The edge of the equilibrium shape (in white) is thus quite wide but can be superposed to the one (in black) calculated as the pedal of the beta-plot $\beta(\theta)=\beta_0+\beta_1\cos3\theta$ (with $\beta_0$=4.12 and $\beta_1$=0.31) depicted in figure 2

**Figure 1**

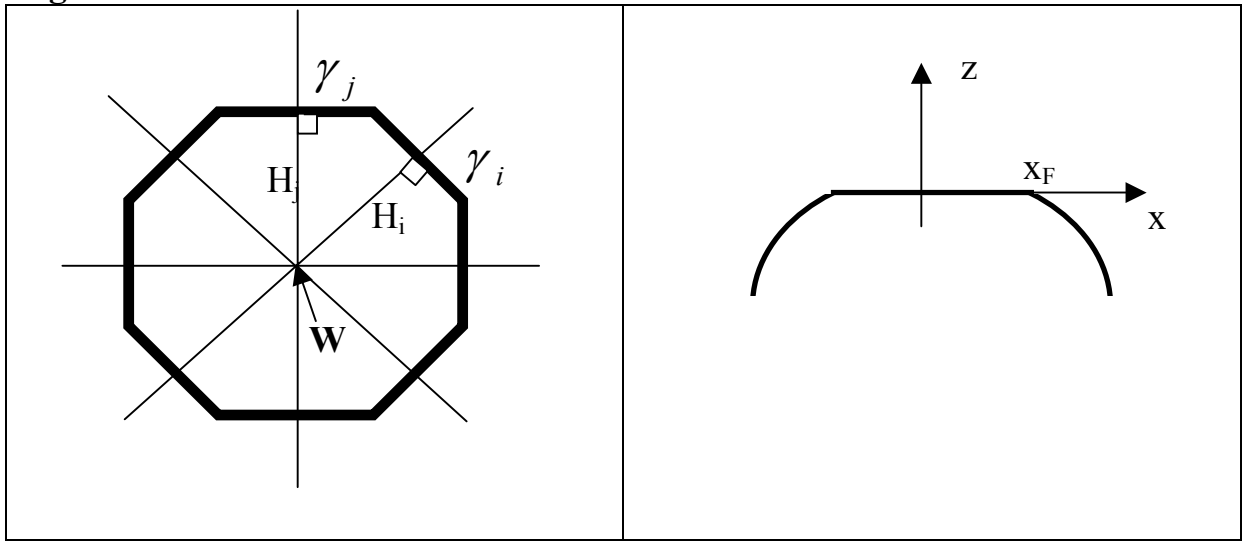

**Figure 2 :**

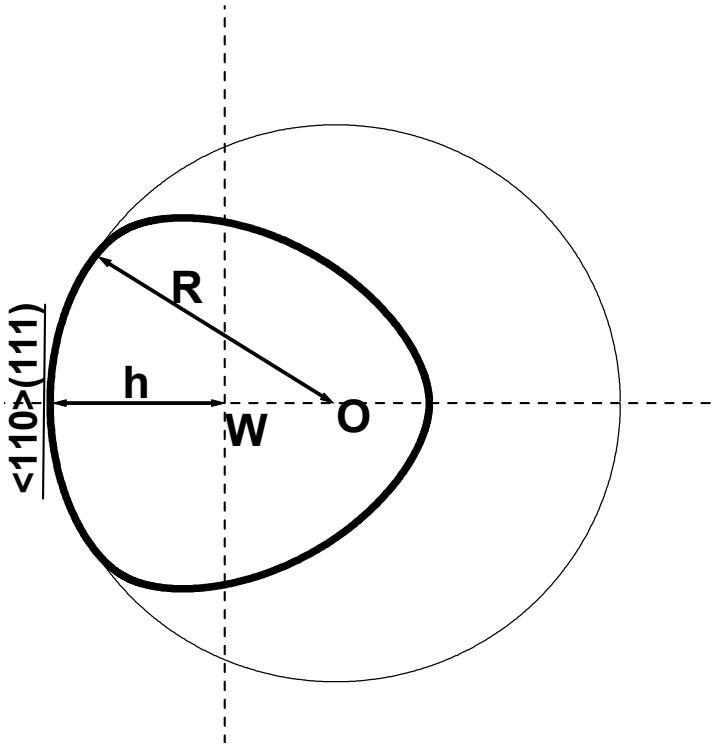

**Figure 3 :**

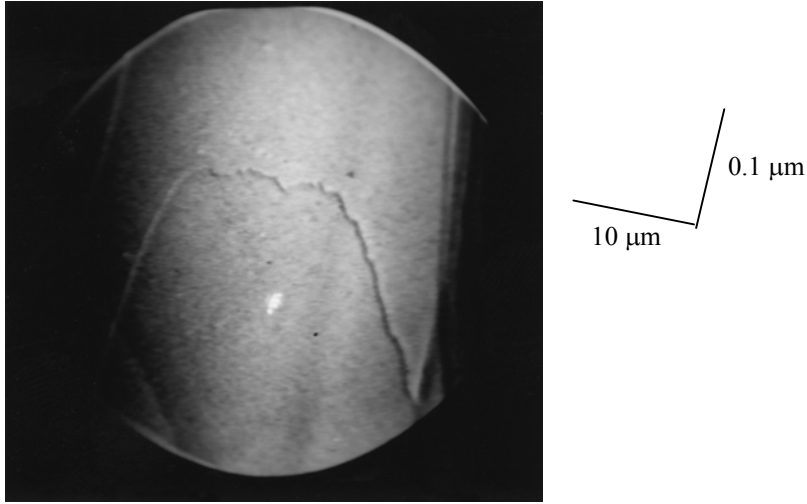

Figure 4

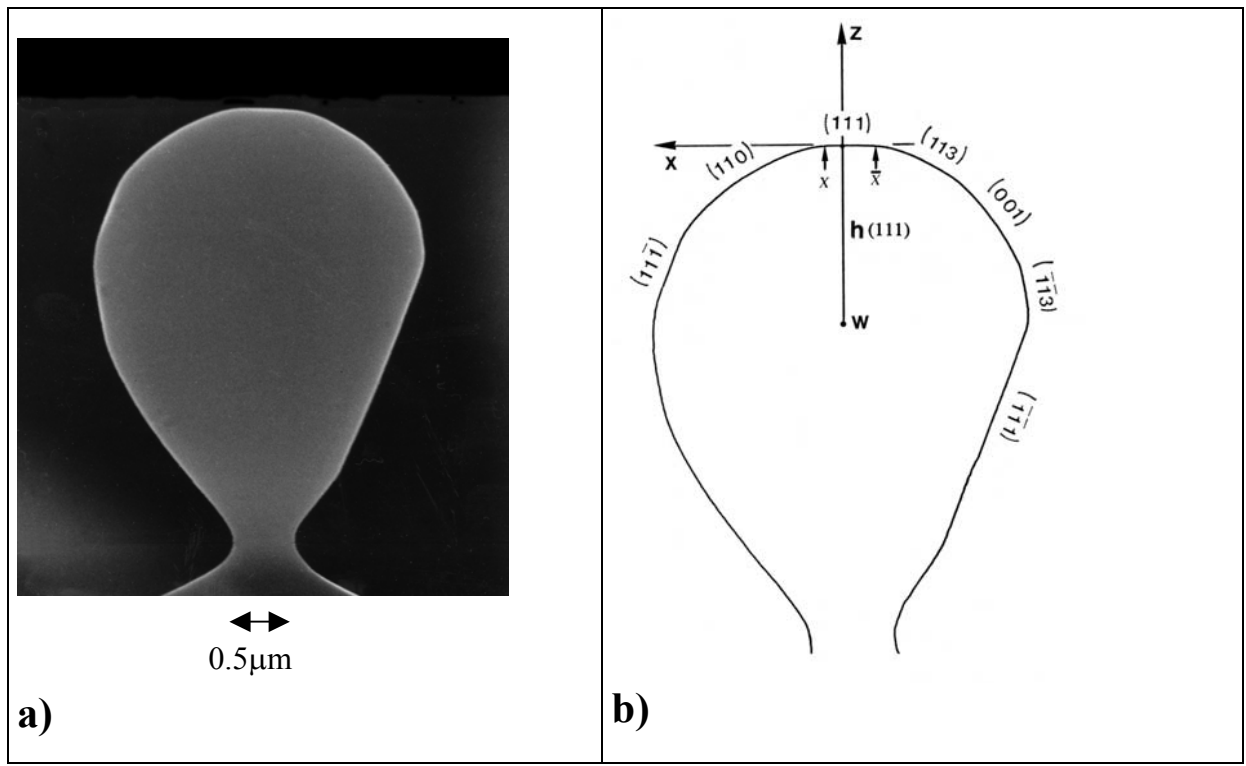

a)             b)

**Figure 5 :**

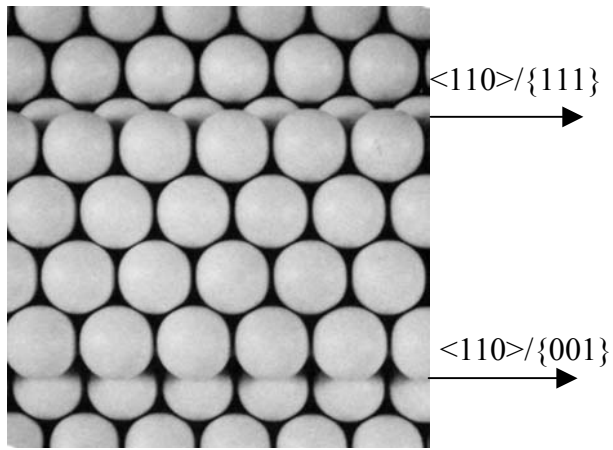

**Figure 6 :**

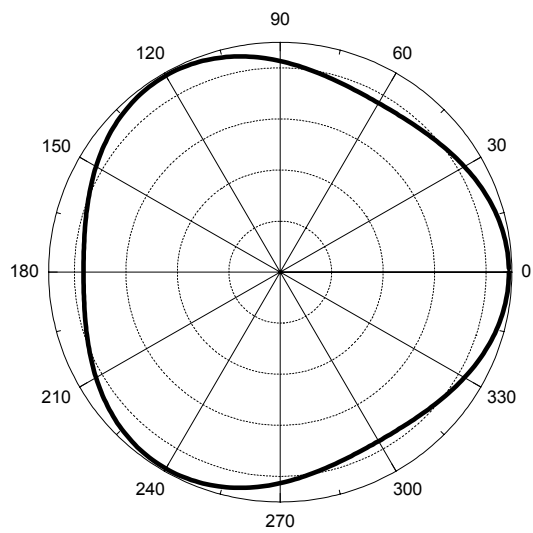

**Figure 7**

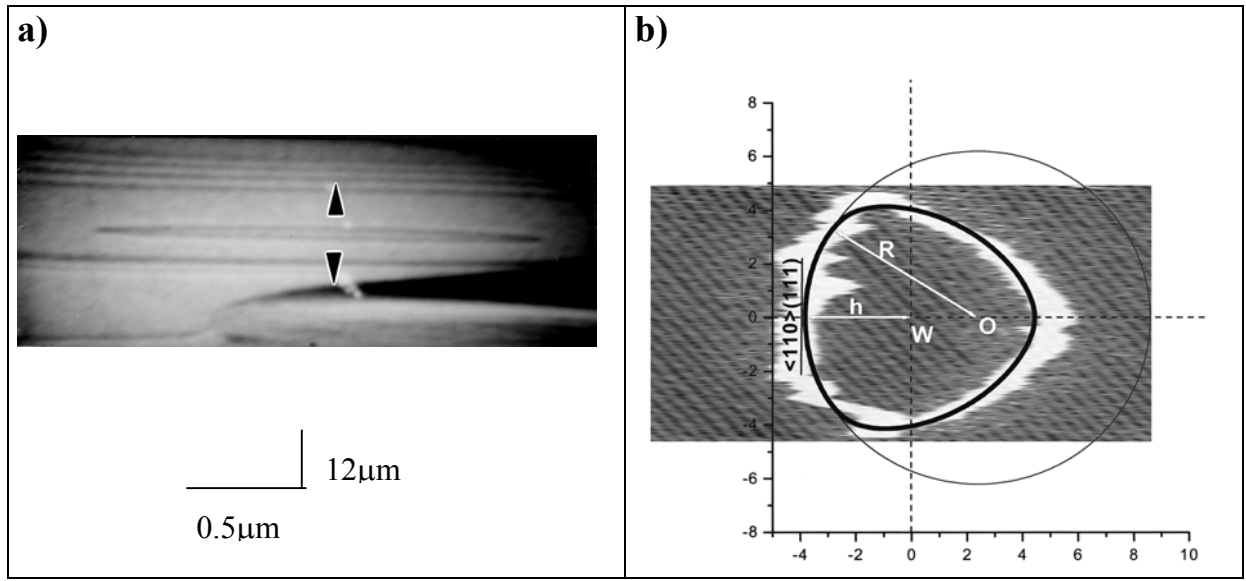